\documentclass[]{aa}  

\usepackage{graphicx}
\usepackage{subfig}
\usepackage{txfonts}
\usepackage{graphicx,rotating}
\usepackage{subfig}
\usepackage{txfonts}
\usepackage[T1]{fontenc}
\usepackage{epsfig}
\usepackage{natbib}
\usepackage{color}
\usepackage{natbib}
\usepackage{amssymb}  
\usepackage{tabularx}
\usepackage{comment}
\DeclareGraphicsExtensions{.pdf,.png,.jpg}
\usepackage[nomarkers,nolists]{endfloat}
\bibpunct{(}{)}{;}{a}{}{,} 
\begin{document}

   \title{Molecular gas in two companion cluster galaxies at $z=1.2$}

   \author{G. Castignani
          \inst{1,2}\fnmsep\thanks{e-mail: gianluca.castignani@obspm.fr}
          \and
          F. Combes\inst{1,2}
          \and
          P. Salom\'e\inst{1}
          \and
          S. Andreon\inst{3}
          \and
          M. Pannella\inst{4} 
          \and
          I. Heywood\inst{5,6} 
          \and 
          G. Trinchieri\inst{3}
          \and
          C. Cicone\inst{3}
          \and
          L.~J.~M. Davies\inst{7}
          \and 
          F.~N. Owen\inst{8} 
          \and
          A. Raichoor\inst{9}           
          }

   \institute{Sorbonne Universit\'{e}, Observatoire de Paris, Universit\'{e} PSL, CNRS, LERMA, F-75014, Paris, France
               \and
             Coll\`{e}ge de France, 11 Place Marcelin Berthelot, 75231 Paris, France
             \and
             INAF-Osservatorio Astronomico di Brera, via Brera 28, 20121 Milano, Italy
             \and
             Department of Physics, Ludwig-Maximilians-Universit\"{a}t, Scheinerstr. 1, D-81679 M\"{u}nchen, Germany
             \and
             Astrophysics, Department of Physics, University of Oxford, Keble Road, Oxford OX1 3RH, UK
             \and
             Department of Physics and Electronics, Rhodes University, PO Box 94, Grahamstown, 6140, South Africa
             \and
             International Centre for Radio Astronomy Research, University of Western Australia M468, 35 Stirling Highway, Crawley, WA 6009, Australia
             \and
             National Radio Astronomy Observatory, P.O. Box O, Socorro, NM 87801, USA
             \and
             Institute of Physics, Laboratory of Astrophysics, Ecole Polytechnique F\'ed\'erale de Lausanne (EPFL), Observatoire de Sauverny, 1290 Versoix, Switzerland
               }

                \date{Received 23 February 2018 / Accepted 5 June 2018}
 
  \abstract
   {Probing both star formation history and evolution of distant cluster galaxies is essential to evaluate the effect of dense environment on shaping the galaxy properties we observe today.}
   {We investigate the effect of cluster environment on the processing of the molecular gas in distant cluster galaxies.
   We study the molecular gas properties of two star-forming galaxies separated by 6~kpc in the projected space and belonging to a galaxy cluster selected from the Irac Shallow Cluster Survey, at a redshift $z=1.2$, that is, $\sim2$~Gyr after the cosmic star formation density peak.  This work describes the  first CO detection from $1<z<1.4$ star-forming cluster galaxies with no {clear reported} evidence of active galactic nuclei.}
   {We exploit observations taken with the NOEMA interferometer at $\sim3$~mm to detect CO(2-1) line emission from the two selected galaxies, unresolved by our observations.}
   {Based on the CO(2-1) spectrum, we estimate a total molecular gas mass $M({\rm H_2})=(2.2^{+0.5}_{-0.4})\times10^{10}$~$M_\odot$, where fully excited gas is assumed, and a dust mass $M_{\rm dust}<4.2\times10^8~M_\odot$ for the two blended sources. The two galaxies have similar stellar masses and H$\alpha$-based star formation rates (SFRs) found in previous work, as well as a large relative velocity of $\sim$400~km/s estimated from the CO(2-1) line width. These findings tend to privilege a scenario where both sources contribute to the observed CO(2-1).
   {Using the archival Spitzer MIPS flux at 24~$\mu$m we estimate an ${\rm SFR(24\mu m)}=(28^{+12}_{-8})~M_\odot$/yr for each of the two galaxies.}
   Assuming that the two sources contribute equally to the observed CO(2-1), our analysis yields a depletion timescale of ${\tau_{\rm dep}=(3.9^{+1.4}_{-1.8})\times10^8}$~yr, and  a {molecular gas to stellar mass ratio of} ${0.17\pm0.13}$ for each of two sources, separately.    
   We also provide a new, more precise measurement of an unknown weighted mean of the redshifts of the two galaxies, $z=1.163\pm0.001$.}
   {{ Our results are in overall agreement with those of other distant cluster galaxies and with model predictions for main sequence (MS) field galaxies at similar redshifts. The two target galaxies have molecular gas mass and depletion times that are marginally compatible with, but smaller than those of MS field galaxies, suggesting that the molecular gas has not been sufficiently refueled. We speculate that the cluster environment might have played a role in preventing the refueling via environmental mechanisms such as galaxy harassment, strangulation, ram-pressure, or tidal stripping.} Higher-resolution and higher-frequency observations will enable us to spatially resolve the two sources and possibly distinguish between different gas processing mechanisms. }


   \maketitle
%

\section{Introduction}\label{sec:introduction}
Evolution and growth of galaxies in clusters are believed to be strongly affected by their dense megaparsec-scale environments.
In the local Universe, we observe tight correlations between cluster occupancy and galaxy properties such as morphology \citep{Dressler1980}, color \citep{Kodama2001}, stellar mass \citep{Ostriker_Tremaine1975}, active galactic nuclei (AGN) fraction \citep{Kauffmann2004}, star formation \citep{Butcher_Oemler1984,Peng2010} and gas content \citep{Gunn_Gott1972,Chung2009,Vollmer2012,Jachym2014}.

There is still debate on the astrophysical processes occurring in clusters which lead to these relations, and the epochs during which they are being established. For example, the cores of nearby rich clusters are largely populated by passive ellipticals, with negligible ongoing star formation \citep[morphology vs. density, and star formation rate vs. density relations,][]{Andreon2006,Raichoor_Andreon2012}, which to first order is consistent with a burst of star formation at $z\gtrsim2$, followed by passive evolution to the present day \citep[e.g.,][]{Eisenhardt2008,Mei2009,Mancone2010}. However, a number of recent studies have found potentially conflicting results to this somewhat simplistic hypothesis, finding a high concentration of potentially in-falling, star-forming galaxies in the outskirts of nearby clusters \citep[e.g.,][]{Bai2009,Chung2010}, and a strong evolution of the fraction of star-forming galaxies in cluster cores out to $z\sim1$ and beyond \citep{Smith2010,Webb2013,Brodwin2013,Zeimann2013,Alberts2016}. These findings appear to contradict the predictions of early, bursty star formation in cluster core galaxies and suggest the late assembly of cluster members via both strong environmental quenching mechanisms 
\citep[e.g., strangulation, ram pressure stripping, and galaxy harassment;][]{Larson1980,Moore1999} and rapid infall of gas feeding star formation at high-$z$ (reaching a maximum at $z\sim2$), followed by slow cooling flows at low-$z$ \citep{Salome2006,Salome2008,Salome2011,Ocvirk2008,Dekel2009a,Dekel2009b}. 


Probing molecular gas in distant clusters is a unique tool to study the star formation history and evolution of cluster galaxies. Advancements of observational facilities at millimeter wavelengths, such as NOEMA and ALMA interferometers, have remarkably increased the statistics of distant $z>1$ cluster galaxies with detections of molecular gas. 
Several studies have investigated molecular gas reservoirs of distant $z>1$ cluster galaxies \citep[see e.g.,][and references therein]{Noble2017}. However, to the best of our knowledge, the cluster galaxy at $z=1.1147$ of \citet{Wagg2012}, classified by the authors as an obscured AGN from the optical spectrum, is the only one in the literature in the redshift range $1<z<1.4$ with a reported CO detection. 

All remaining studies refer to cluster galaxies in the range ${z\sim1.4-2.0}$, that is, close to the peak of star formation activity in clusters occurring at $z\sim2$. 
{ We summarize in the following the main results from the literature in this specific redshift range. 
\citet{Aravena2012} detected CO(1-0) in two galaxies belonging to a cluster candidate at $z\simeq1.55$. 
\citet{Casasola2013} detected CO(2-1) in the gas-rich radio galaxy 7C~1756+6520 in   cluster environment at $z\simeq1.4$.
\citet{Noble2017} reported the detection of CO(2-1) in 11 gas-rich galaxies belonging to three galaxy clusters at $z\simeq1.6$.
\citet{Rudnick2017} detected CO(1-0) in two galaxies from a confirmed galaxy cluster at $z\simeq1.6$.
\citet{Webb2017} reported the discovery of a large molecular  gas reservoir based on CO(2-1) observations of a distant brightest cluster galaxy at $z=1.7$.
\citet{Stach2017} and \citet{Hayashi2017,Hayashi2018} detected molecular gas in a sample of 17 galaxies in total, belonging to the cluster XMMXCS~J2215.9-1738 at $z\simeq1.5$.}
{ Recently, \citet{Coogan2018} detected molecular gas in a sample of eight cluster galaxies belonging to the distant cluster Cl~J1449+0856  at $z=1.99$.}

At even higher redshifts, that is, $z\gtrsim2$, large reservoirs of molecular gas have commonly been observed in highly star-forming galaxies in proto-clusters, with star formation rates SFR$\gtrsim200~M_\odot$/yr  \citep{Papadopoulos2000,Emonts2016,Wang2016,Oteo2018}.

At lower redshifts, $1<z<1.4$, the molecular gas content of distant cluster galaxies is still substantially unexplored. Such an epoch is nevertheless important for understanding both the fate of molecular gas in distant cluster galaxies and the mechanisms responsible for the growth of cluster galaxies, as $z\gtrsim2$ proto-clusters evolve into virialized structures down to $z\sim0$. 

In the present work, we report the detection of CO(2-1) from two unresolved cluster galaxies at $z=1.2$ obtained with the NOEMA interferometer at the Plateau de Bure. This represents the first CO detection from $1<z<1.4$ cluster galaxies with no reported { clear} evidence of AGNs.  {  Our two selected galaxies are observed significantly later ($\sim2$~Gyr) than the cosmic star formation density peak occurring at $z\sim2$ for field galaxies \citep[e.g.,][]{Madau_Dickinson2014}, and possibly earlier for (proto-)cluster galaxies \citep{Chiang2017,Overzier2016}.}  The two sources have an angular separation of 0.72~arcsec inferred from their archival Hubble Space Telescope (HST) image (i.e., 6~kpc at the redshift of the sources) and belong to the distant cluster ISCS~J1426.5+3339. They are located at an angular separation of $\sim$1~arcmin ($\sim$500~kpc) from the cluster center. The CO detection reported in this work is the first one resulting from a wider search for molecular gas in distant cluster galaxies.

In Sect.~\ref{sec:sample}, we describe the two targets; in Sect.~\ref{sec:observations_and_data_reduction} we describe the observations and data reduction; in Sects.~\ref{sec:results} and \ref{sec:discussion} we present and discuss the results, respectively; in Sect.~\ref{sec:conclusions} we draw our conclusions.

Throughout this work, we adopt a flat $\Lambda \rm CDM$ cosmology with matter density $\Omega_{\rm m} = 0.30$, dark energy density $\Omega_{\Lambda} = 0.70$ and Hubble constant $h=H_0/100\, \rm km\,s^{-1}\,Mpc^{-1} = 0.70$ \citep[see however,][]{PlanckCollaborationXIII2016,Riess2016,Riess2018}.  Under these assumptions, the luminosity distance of the target sources is $D_L=8.024$~Gpc.

\section{Two cluster galaxies}\label{sec:sample}
We consider the sample of 18 galaxy clusters at $1.0<z<1.5$ from the \citet{Zeimann2013} catalog. The clusters belong to the Irac Shallow Cluster Survey (ISCS) that comprises distant clusters detected within the  Bo\"{o}tes field of the NOAO Deep Wide-Field Survey \citep[NDWFS;][]{Jannuzi_Dey1999} using accurate optical/infrared photometric redshifts \citep{Brodwin2006} and the 4.5~$\mu$m flux-limited catalog of the IRAC Shallow Survey \citep[ISS;][]{Eisenhardt2004}. 
Mass estimates were obtained for a subset of the ISCS clusters using X-ray emission, weak lensing measurements, or dynamical arguments \citep{Brodwin2011,Jee2011,Andreon2011}, as well as clustering analysis of the full ISCS sample \citep{Brodwin2007}.

Our goal is to observe actively star-forming cluster galaxies and to explore the future fate of these populations. Therefore, we only select spectroscopically confirmed cluster members from the \citet{Zeimann2013} catalog with the highest H$\alpha$-based SFRs (i.e., $\gtrsim 120 M_\odot$/yr). This selection yields two galaxies with spectroscopic redshift of $z=1.17$ belonging to the cluster ISCS~J1426.5+3339 \citep{Zeimann2013}. The halo mass of ISCS~J1426.5+3339 is expected to be in the range  $\sim(0.8-2)\times10^{14}~M_\odot$, corresponding to a virial radius of $\sim1$~Mpc \citep{Wagner2015}.
Each of the two selected galaxies has H$\alpha$-based SFR$\simeq130~M_\odot$/yr.  In Table~\ref{tab:cluster_galaxies_properties} we report the properties of the two targets. 

Neither of the targets is a plausible AGN candidate based on both infrared diagnostic and X-ray emission \citep{Zeimann2013}, as described in the following. The cluster ISCS~J1426.5+3339 falls in fact within the X-ray XBo\"{o}tes survey region \citep{Kenter2005,Murray2005} and has deep Chandra observations \citep{Martini2013} that enable the characterization of the AGN population of the cluster. AGN candidates within the sample of cluster galaxies of \citet{Zeimann2013} were identified and distinguished by the authors from star-forming galaxies either by means of their X-ray luminosities, or by adopting the empirical mid-infrared criteria by \citet{Stern2005}. Sources matched to the Spitzer Deep Wide-Field Survey \citep[SDWFS;][]{Ashby2009} catalogs with signal-to-noise ratio S/N$>5$ in all four IRAC bands that fall in the AGN edge of the  diagnostic WISE color-color diagram by \citet{Stern2005} were also conservatively classified as AGNs by \citet{Zeimann2013}.

We also checked that the two sources considered in this work are not included in the catalog of the Very Large Array Faint Images of the Radio Sky at Twenty-Centimeters (VLA FIRST) survey at 1.4~GHz \citep{Becker1995}. The detection limit of the FIRST catalog is $\sim$1~mJy (with a typical root mean square (rms) of 0.15~mJy). Post-pipeline radio maps have a resolution of $\sim$5 arcsec.  The two selected sources are therefore unresolved by VLA FIRST. Assuming the flux value of $\sim$1~mJy as upper limit, we estimate an isotropic rest frame 1.4~GHz luminosity density $L_\nu<7\times10^{31}$~erg~s$^{-1}$~Hz$^{-1}$ associated with the blended emission from the two sources. A fiducial spectral index $\alpha=0.8$ and the power-law $L_\nu\propto\nu^{-\alpha}$ are assumed for the calculation. The estimated upper limit implies that if the two target sources host radio galaxies, they have low non-thermal radio luminosities, compatible with galaxies populating the faint end of the radio luminosity function.


\section{Observations and data reduction}\label{sec:observations_and_data_reduction}
\begin{figure}[]\centering
\vspace{-.2cm}
\subfloat{\includegraphics[width=0.8\textwidth]{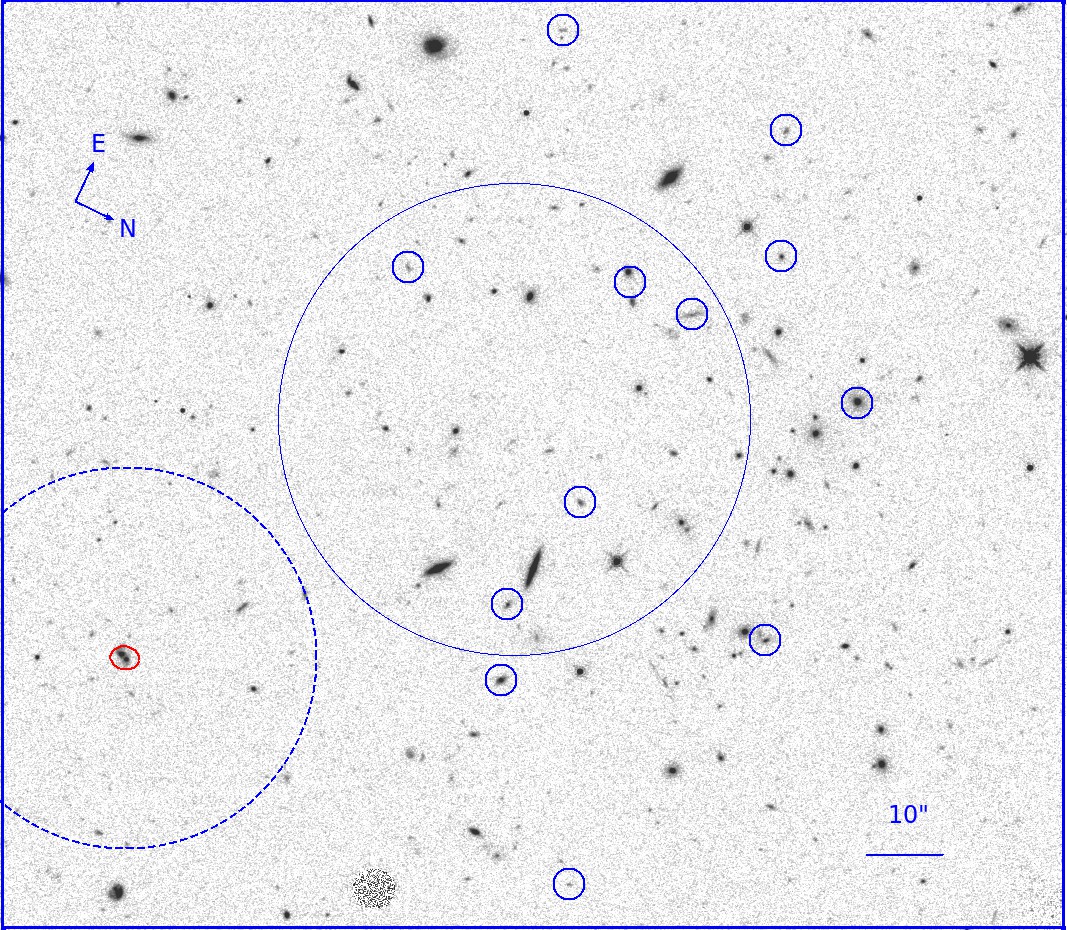}}\\
\subfloat{\includegraphics[width=0.4\textwidth]{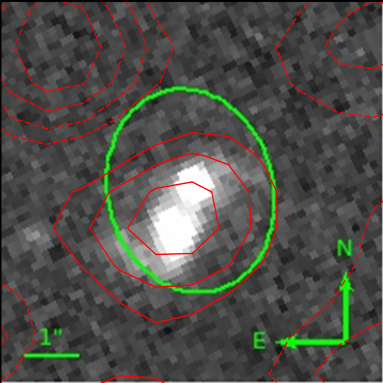}}
\caption{Top: archival HST WFC3 image (${\rm width}=136''$; ${\rm height}=119''$) of the cluster ISCS~J1426.5+3339, taken with the F160W filter. The central, big,  blue, and solid circle has a radius of 250~kpc. The NOEMA beam size with our two targets (red ellipse at the bottom left) and the NOEMA field of view (dashed blue arc) are shown. The small blue circles denote additional spectroscopically confirmed cluster members \citep{Eisenhardt2008,Zeimann2013}. 
Bottom: $7''\times7''$ zoom image showing the galaxy J142626.1+333827 and its southern companion J142626.1+333826. The green ellipse shows the NOEMA beam size (major axis $ = 3.69''$; minor axis $ = 2.98''$). Red contour levels correspond to the CO(2-1) detection, as in Fig.~\ref{fig:NOEMA_observations}, left panel.}
\label{fig:HST_image}
\end{figure}

We observed the two sources with eight antennas using the compact array D-configuration of the NOEMA interferometer (P.I.:~Castignani). Such a configuration  provides the lowest phase noise and the highest surface brightness sensitivity. It is therefore best suited for detection experiments such as ours.
Our observations aim at detecting CO(2-1) from the two sources at observer frame frequency $\nu_{\rm obs}=106.239$~GHz (i.e., $\nu_{\rm rf}=230.538$~GHz in the rest frame), given the spectroscopic redshift $z=1.17$ \citep{Zeimann2013}.  At an angular resolution of $\sim3$~arcsec, the two target sources are unresolved. The phase center of our observation is R.A.~=~14h:26m:26.090s, Dec.~=~33d:38m:27.600s, corresponding to the J2000 coordinates of the northernmost galaxy among the two \citep{Zeimann2013}. Figure~\ref{fig:HST_image} shows the HST images, taken with the WFC3/infrared camera \citep{Kimble2008}, of the two unresolved sources along with the NOEMA beam (bottom), as well as of the cluster field (top). 
{ The image at the top is centered at the cluster center coordinates, corresponding to the galaxy overdensity peak found using photometric redshifts of galaxies \citep{Eisenhardt2008}.} We employed the WideX correlator which provides 3.6~GHz of instantaneous dual-polarization bandwidth.
The observations were carried out over four days (15~Jul, 29~Jul, 3~Aug, and 6 Aug, 2017) in good to average weather conditions: { radio seeing  at ${\nu_{\rm obs}}$}  in the range $\sim(1.43-1.81)$~arcsec and  precipitation water vapor in the range $\sim(3-15)$~mm. { The on-source observing time is $\sim$7.7~hr (12.3~hr in total, including overheads)}, corresponding to 987 scans. 
Data reduction was performed using the latest release of GILDAS package as of May~2017\footnote{https://www.iram.fr/IRAMFR/GILDAS/}. Data were calibrated using standard pipeline adopting a natural weighting scheme to maximize sensitivity. Only minor flagging was required: 27 scans ($\lesssim3\%$ of the total) were removed because of bad weather conditions during their acquisition. 


\section{Results}\label{sec:results}
\subsection{Significance of the detection}
\begin{figure}[]\centering
\subfloat{\includegraphics[width=0.5\textwidth]{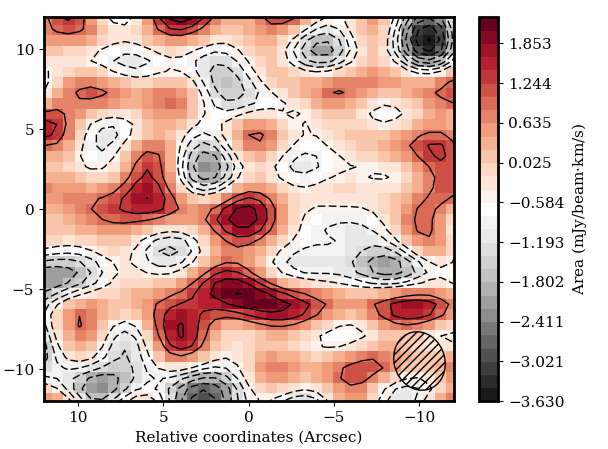}}
\subfloat{\includegraphics[width=0.6\textwidth]{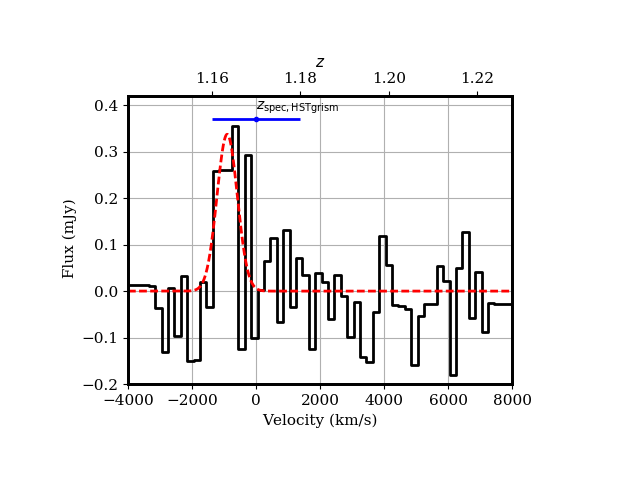}}
\caption{Left: intensity map, cleaned and corrected for the NOEMA primary beam showing the detection of the two unresolved target sources at the center. Coordinates are reported as angular separations from the source J142626.1+333827. The velocity range considered is within -2,000~km/s and 0~km/s, corresponding to the CO(2-1) line, see right panel.
Solid and dashed contour levels are superimposed and correspond to positive and negative fluxes, respectively. The dashed ellipse (bottom right) shows the beam size. Right: spectrum  (black solid line) obtained with NOEMA within an aperture corresponding to the beam size at the location of the detection, i.e., the center of the intensity map. A resolution of 200~km/s is adopted. The best fit, baseline subtracted, is reported (red dashed line).  
The center and the length of the blue horizontal segment at the top of the spectrum correspond to the HST WFC3 spectroscopic redshift and its uncertainty, respectively \citep{Zeimann2013}.}
\label{fig:NOEMA_observations}
\vspace{-.1cm}
\includegraphics[width=0.5\textwidth]{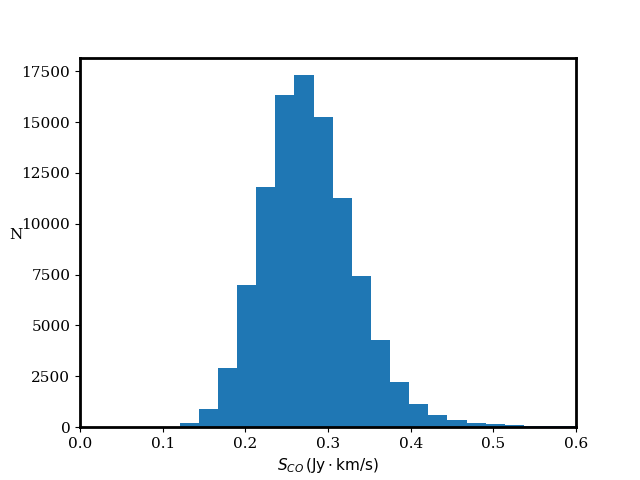}
\caption{Distribution of the integrated CO(2-1) fluxes, as derived from our Monte Carlo simulations.}
\label{fig:SCO_dist}
\end{figure}

In Fig.~\ref{fig:NOEMA_observations} we present the results of our NOEMA observations where the detection of the two blended sources from the intensity map and the corresponding CO(2-1) spectrum at a resolution of 200~km/s are shown. { Consecutive levels in the intensity map correspond to an increment and a decrement of 0.5~rms, starting from the values of +1~rms and -1~rms, respectively. The rms=0.9~mJy/beam$\cdot$km/s is estimated here from the map, considering the region defined by relative coordinates, with respect to those of J142626.1+333827,   with absolute values in the range between 4 and 12~arcsec.

We fit the spectrum} using the $\chi^2$ minimization procedure with a best-fit model given by the sum of a Gaussian and a polynomial of degree one to account for the CO(2-1) spectral line and the baseline, respectively, { as shown in Fig.~\ref{fig:NOEMA_observations} (right panel).} The fit yields $\chi^2/{\rm d.o.f.}=1.39$, where ${\rm d.o.f.}$ is the number of degrees of freedom with five best fit parameters, that is,  the intercept and the slope of the polynomial describing the baseline, as well as the normalization, the width, and the mean of the Gaussian. The fit yields a CO(2-1) peak flux of 0.34~mJy. The best fit is reported in the right panel of Fig.~\ref{fig:NOEMA_observations}. 


At 106.239~GHz, the field of view of NOEMA is $\sim50$~arcsec (25~arcsec radius). We visually inspected both the archival HST image (Fig.~\ref{fig:HST_image}) and the radio intensity map (Fig.~\ref{fig:NOEMA_observations}) within the NOEMA field of view. No serendipitous detection of additional cluster, background, or foreground sources has been found.

To estimate both the significance of the detection and the uncertainties in the best fit parameters, we adopt criteria based on $\chi^2$ statistics and perform Monte Carlo simulations, as described in the following. We derive the rms dispersion around the mean of the spectrum within the velocity range between 1,000~km/s and 7,000~km/s; this gives an estimate of the noise level equal to 79~$\mu$Jy. 

We then generate $N=100,000$ simulated spectra assuming that at fixed velocity channel the flux value of each simulated spectrum is drawn from a Gaussian distribution with a mean equal to the observed flux at that velocity channel and a variance equal to the square of the estimated noise level. Then we fit each simulated spectrum with the $\chi^2$ minimization procedure  as described above. The resulting $N$ fits are reasonably good, with $\chi^2/{\rm d.o.f.}<4$ for all of them.

We estimate the expected value as well as both the upper and lower uncertainties of each best fit parameter, as the median value of the corresponding distribution, and the values delimiting both the upper and lower 15.865\% quartiles of the distribution, respectively.
The statistical errors associated with the peak velocity and the width of the CO(2-1) line are estimated also with standard criteria based on $\chi^2$ statistics \citep{Cash1976} applied to the observed spectrum. 
The estimated uncertainties are $\sim$70~km/s, lower than the adopted channel width. Therefore, we consider the latter as fiducial uncertainty for both best fit parameters. 

The resulting estimated full width at half maximum (FWHM) of the detected CO(2-1) line is $(733\pm200)$~km/s, which is large when compared to the typical width of $\sim$200-300~km/s. We refer to Sect.~\ref{sec:discussion} for a discussion. The best fit peak velocity of the CO(2-1) line is translated into a redshift estimate $z$ for the two unresolved sources.
Our analysis yields $z=1.163\pm0.001$, fully consistent within the uncertainties with that reported by \citet{Zeimann2013} using HST grism spectroscopy (see Table~\ref{tab:cluster_galaxies_properties}).

To assess the significance of the detection we estimate the integrated CO(2-1) fluxes using the Gaussian best fit model of each simulated spectrum. This procedure implies a velocity integrated CO(2-1) flux  $S_{CO(2-1)}\Delta \varv=(0.27^{+0.06}_{-0.05})$~Jy~km/s, where the reported value and errors are the median and the uncertainties within the $68.27\%$ confidence region of the distribution of the integrated CO fluxes, as described above. The distribution is plotted in Fig.~\ref{fig:SCO_dist}. The estimated flux and uncertainty imply that our detection has a ${\rm S/N}=5.4$. 
{ As a consistency check, we note that}  the mean values of both estimated flux and redshift found with Monte Carlo simulations are equal to those found from the best fit model of the observed spectrum. The fact that the two unresolved galaxies are detected in CO(2-1) by NOEMA both in projected space at the source coordinates and in frequency at the source redshift further supports the reliability of both our detection and the S/N estimate. 

{ We stress that the noise associated with interferometric observations is not Gaussian in the projected space. In fact, spurious peaks due to the side lobes can occur at coordinates that are not coincident with the phase center of our observations, also considering the low S/N of the our detection. These aspects prevent us from estimating the S/N directly from the intensity map of Fig.~\ref{fig:NOEMA_observations}a. Instead, similarly to previous work \citep{Walter2016,Decarli2016}, we need to rely on a positional prior: we cannot search for CO detections within the entire data cube (rms=0.3~mJy/beam at 100~km/s resolution) without relying on robust HST counterparts. Our target galaxies are in fact clearly detected  \citep[based on HST observations,][]{Zeimann2013} in both redshift and projected coordinates, independently of our CO observations.}

\subsection{Molecular gas mass}
We derive the CO(2-1) luminosity $L^{\prime}_{CO(2-1)}$ for the two blended sources from the velocity integrated CO(2-1) line flux using Eq.~(3) of \citet{Solomon_VandenBout2005}:
\begin{equation}
\label{eq:LpCO}
 L^{\prime}_{CO(2-1)}=3.25\times10^7\,S_{CO(2-1)}\,\Delta\varv\,\nu_{\rm obs}^{-2}\,D_L^2\,(1+z)^{-3}\,.
\end{equation}
We obtain $L^{\prime}_{CO(2-1)}=(4.9^{+1.1}_{-0.9})\times10^{9}$~pc$^2$~K~km~s$^{-1}$, typical of galaxies with infrared luminosities $\gtrsim10^{11}~L_\odot$; see, for example, Fig.~8 of \citet[][]{Solomon_VandenBout2005}.
Using $L^{\prime}_{CO(2-1)}$ and standard relations we estimate a number of additional physical properties, as described in the following.

By assuming a Galactic CO-to-H$_2$ conversion factor $X_{CO}\simeq2\times10^{20}$~cm$^{-1}$/(K km/s), that is, $\alpha_{CO}=4.36~M_\odot\,({\rm K~km~s}^{-1}~{\rm pc}^2)^{-1}$, typical of main sequence (MS) galaxies \citep{Solomon1997,Bolatto2013}, we estimate a total molecular gas mass of $M({\rm H_2})=\alpha_{CO}L^{\prime}_{CO(2-1)}=(2.2^{+0.5}_{-0.4})\times10^{10}~M_\odot$ for the two blended sources. We note that the conversion factor $\alpha_{CO}$ is commonly used to convert $L^{\prime}_{CO(1-0)}$ into $M({\rm H_2})$, where $L^{\prime}_{CO(1-0)}$ is the velocity integrated CO(1-0) luminosity. Therefore we implicitly assume fully excited gas in our calculation, that is, $r_{21}= L^{\prime}_{CO(2-1)}/L^{\prime}_{CO(1-0)}=1$, { typical of distant star-forming galaxies, or at least a fraction of them \citep[][see also Sect.~\ref{sec:discussion} for a discussion]{Daddi2015}}.

We stress here that we do not know the relative contribution of each of the two unresolved sources to the total CO(2-1) flux. Therefore we consider in the following the scenarios in which the observed flux is either due to one of the two sources or to both of them. Since the two galaxies have similar H$\alpha$-based SFRs and equal stellar masses estimated by \citet{Zeimann2013} we will not be able to distinguish which of the two sources contributes more significantly to the observed CO(2-1) flux. We define as $f$ the fraction of the observed CO(2-1) flux that is due to J142626.1+333827. Given such a notation, the two cases $f=0$ and $f=1$ are considered as indistinguishable. We consider in the following, when appropriate to estimate physical properties, the two extreme cases where $f=0$ and $f=1/2$, keeping in mind that the correct values will be intermediate between the two. Physical quantities related to $f=1/2$ refer to either J142626.1+333827 or J142626.1+333826, considered individually. On the other hand, those related to $f=0$ refer to only one of the two galaxies, assuming that this source { contributes the total} CO(2-1) flux. 

\subsection{Star formation rate}
{ We exploit our observations to estimate the SFR independently from \citet{Zeimann2013} by using Spitzer MIPS observations in the mid-infrared. We search for our sources in the Spitzer Enhanced Imaging Products (SEIP) source catalog\footnote{https://irsa.ipac.caltech.edu/data/SPITZER/Enhanced/SEIP/ overview.html}.   
A Spitzer MIPS detection at 24~$\mu$m in the observer frame is found at an angular separation of 1.44 and 0.72~arcsec from J142626.1+333826 and J142626.1+333827, respectively.  Our target sources are unresolved by Spitzer MIPS, given its angular resolution of $\sim$6~arcsec at 24~$\mu$m, corresponding to the FWHM of the telescope point spread function. The 24~$\mu$m flux of the Spitzer MIPS source is $(2.755\pm0.122)\times10^2\mu$Jy that we use to estimate the SFR. We use a procedure analogous to that adopted by \citet{McDonald2016}. We estimate the $\lambda=24~\mu$m rest frame luminosity $\lambda L_\lambda$ by assuming a power-law model, $L_\lambda\propto\lambda^\gamma$, with $\gamma=2.0\pm0.5$ \citep{Casey2012}. We incorporate the uncertainties in both $\gamma$ and the observed 24~$\mu$m flux by using $M=100,000$ values drawn from Gaussian distributions centered at the mean values and with standard deviations equal to the reported uncertainties. Then we use the \citet{Calzetti2007} relation to convert the  24~$\mu$m rest frame luminosity into an estimate of the SFR. This procedure yields ${\rm SFR(24\mu{\rm m})}=(52^{+21}_{-15})~M_\odot$/yr ($f=0$) and ${\rm SFR(24\mu{\rm m})}=(28^{+12}_{-8})~M_\odot$/yr ($f=1/2$). The reported SFR values and uncertainties are the medians and the 68.27\% confidence levels estimated from the $M$ realizations, respectively. The notation ($24\mu{\rm m}$) refers to the fact that the SFR is estimated using 24~$\mu$m fluxes.} 

Taking the uncertainties in the SFR estimates into account, the ${\rm SFR(24\mu{\rm m})}$ is marginally consistent with the H$\alpha$-based ${{\rm SFR}=(127^{+100}_{-75})M_\odot/{\rm yr}}$ and ${(134^{+100}_{-75})M_\odot/{\rm yr}}$, estimated by \citet{Zeimann2013} for the galaxies J142626.1+333827 and J142626.1+333826, respectively (see Table~\ref{tab:cluster_galaxies_properties}). Therefore, we suspect that AGN emission might contribute to the observed H$\alpha$ flux, which results in H$\alpha$-based SFRs that are possibly biased-high (see also Sect.~\ref{sec:discussion}). For this reason, we henceforth adopt the ${\rm SFR(24\mu{\rm m})}$ as SFR estimate for our target galaxies. { No evidence for AGN contamination has been found for our target galaxies using mid-infrared criteria by previous work (see Sect.~\ref{sec:sample}). However, we stress here that a possible contamination cannot be excluded at 24~$\mu$m, that is, 11~$\mu$m in the rest frame, where AGN contribution is however typically limited to $\lesssim20\%$ for distant star-forming galaxies \citep{Donley2012,Pozzi2012,Delvecchio2014}.}

{ We also note that the two scenarios of $f=0$ and $f=1/2$ have been introduced for the CO emission. By applying the same $f$-prescription for the estimate of the SFR(24$\mu$m) we implicitly assume that the CO emission correlates well with that at 24$\mu$m. This assumption and our choice for the SFR is strengthened by the fact that the ${\rm SFR(24\mu{\rm m})}$ fully agrees with that  expected from standard $L^\prime_{CO}$ versus SFR relations, such as that found} by \citet{Daddi2010} between the $L^\prime_{CO}$ and the infrared luminosity for both MS and color-selected star-forming galaxies at $z\sim0.5-2.3$. We refer to \citet{Carilli_Walter2013} for a review. By using the relation of \citet{Daddi2010} and  that of \citet{Kennicutt1998a} between the total far-infrared luminosity and the SFR, we obtain   SFR(CO)$=(52^{+33}_{-32})$~$M_\odot/$yr and SFR(CO)$=(24\pm15)$~$M_\odot/$yr for $f=0$ and $f=1/2$, respectively. The reported uncertainties take into account those associated with the estimated CO(2-1) luminosity that are summed in quadrature to fiducial $\sim58\%$, that is, 0.25~dex, uncertainties associated with the calibration and the scatter of the adopted scaling relations \citep{Daddi2010}. The notation (CO) refers to the fact that the SFR is estimated by exploiting CO observations. We note that the relations used to estimate the SFR from the molecular gas properties are ultimately related to the Kennicutt-Schmidt law \citep{Schmidt1959,Kennicutt1998b} which links the molecular gas density to the SFR density, whereas the integrated CO(2-1) flux that is used here to estimate the molecular gas content is not a {\it direct} tracer of star formation.

\subsection{Main sequence and depletion time}
We use the stellar mass ${\rm Log}(M_\star/M_\odot)=10.8\pm0.5$ reported by  \citet{Zeimann2013} for each of two considered galaxies to estimate a number of physical parameters, as described in the following.
Our results imply a specific star formation rate ${{\rm sSFR(24\mu{\rm m})}={\rm SFR(24\mu{\rm m})}/M_\star=(0.82^{+0.65}_{-0.64})}$~Gyr$^{-1}$ and ${(0.44^{+0.36}_{-0.34})}$~Gyr$^{-1}$ in the case of $f=0$ and $f=1/2$, respectively. The reported uncertainties are obtained by taking those associated with both SFR($24\mu{\rm m}$) and $M_\star$ into account.

{  By exploiting the SFR values above, we also estimate a depletion timescale associated with the consumption of the molecular gas, equal to $\tau_{\rm dep}=M({\rm H_2})/{\rm SFR(24~\mu m)}=(4.2^{+1.6}_{-1.9})\times10^8$~yr and $(3.9^{+1.4}_{-1.8})\times10^8$~yr for $f=0$ and  $f=1/2$, respectively.}
{ The reported uncertainties are obtained by propagating and combining in quadrature those associated with the SFR and
$M({\rm H_2})$.} For the sake of clarity, we also derive and report in Table~\ref{tab:cluster_galaxies_properties} the depletion time, as well as the sSFR(H$\alpha$), estimated by adopting SFR(H$\alpha$). The notation (H$\alpha$) refers to the fact that the H$\alpha$-based SFR  is used.
We obtain $\tau_{\rm dep}{\rm (H\alpha)}=M({\rm H_2})/{\rm SFR(H\alpha)}=(0.87^{+0.55}_{-0.70})\times10^8$~yr and $(0.82^{+0.49}_{-0.63})\times10^8$~yr in the case $f=1/2$, for J142626.1+333827 and J142626.1+333826, respectively. 
The reported uncertainties are estimated by summing in quadrature those associated with both SFR(H$\alpha$) and $M({\rm H_2})$. 

Furthermore, we estimate the { molecular gas to stellar mass ratio} $M({\rm H_2})/M_\star$, equal to $0.35^{+0.25}_{-0.26}$ and $0.17\pm{0.13}$ for $f=0$ and $f=1/2$, respectively. The reported uncertainties are estimated by taking those associated with both $M({\rm H_2})$ and $M_\star$ into account. { We refer to Sect.~\ref{sec:discussion} for  a discussion.}

\subsection{Continuum emission and dust mass}
Our observations did not allow us to detect the continuum flux. However, we exploit them to derive an upper limit on the dust mass of the two blended galaxies.

We estimate a 3-$\sigma$ upper limit of 33~$\mu$Jy for the continuum flux $S_{\nu_{\rm obs}}$ of the two blended sources at the observed frequency $\nu_{\rm obs}$, within the 3.6~GHz bandwidth of WideX. { The estimate is consistent with that inferred by assuming a standard dependence of the rms on the inverse of the square root of the bandwidth.}
Following previous work \citep[e.g.,][]{Beelen2006}, the continuum flux can be expressed as a function of the dust mass $M_{\rm dust}$:
\begin{equation}
\label{eq:flux_Mdust}
 S_{\nu_{\rm obs}} = \frac{1+z}{D_L^2}\,M_{\rm dust}\,\kappa(\nu_{\rm rf})\,B_{\nu_{\rm rf}}(\nu_{\rm rf},T_d)\;,
\end{equation}
where $z$ is the redshift of the two selected galaxies, $\kappa(\nu)$ is the dust opacity per unit mass of dust, and $B_{\nu}(\nu,T)$ is the spectral radiance of a black body of temperature $T$ at frequency $\nu$. It holds:\begin{equation}
 B_\nu(\nu,T) = \frac{2h\,\nu^3}{c^2}\frac{1}{e^{\frac{h\nu}{k_BT}}-1}\;,
\end{equation}
where $h$ is the Planck constant, $c$ is the speed of light, and $k_B$ is the Boltzmann constant. 
We assume dust temperature $T_d=25$~K, following recent work by \citet{Scoville2017} on a large sample of $\sim$700 distant galaxies with ALMA detections within the redshift range $z=0.3-4.5$.

We also model the dust opacity as $\kappa(\nu)=\kappa_0(\frac{\nu}{\nu_0})^\beta$, with $\kappa_0=0.4$~cm$^{2}$~g$^{-1}$ , ${\nu_0=\frac{c}{1200\mu{\rm m}}}$ \citep[][and references therein]{Beelen2006,Alton2004}, and $\beta=1.8$ \citep[][]{Scoville2017}. In particular, the value chosen for $\beta$ is based on the determinations for our Galaxy from Planck \citep{PlanckCollaboration2011a,PlanckCollaboration2011b}. We refer to \citet{Berta2016} and \citet{Bianchi2013} for a discussion on possible variations in $\beta$. These assumptions yield a 3-$\sigma$ upper limit for the total dust mass of the two blended sources, $M_{\rm dust}<4.2\times10^8~M_\odot$, consistent with dust reservoirs found for MS galaxies \citep{Berta2016}.

\subsection{Effective radii}
{ We also use SExtractor \citep{Bertin_Arnouts1996} { to estimate the size of our target galaxies} from the HST image reported in Fig.~\ref{fig:HST_image}.  We derive effective radii ${\rm R_e}\simeq(3-4)$~kpc, { where the point spread function has been taken into account by convolving the HST image with a Gaussian filter with a FWHM=2~pixels, corresponding to $\sim2$~kpc at the redshift of our target galaxies.} } The results are summarized in Tables~\ref{tab:cluster_galaxies_properties} and \ref{tab:NOEMAresults}.

\subsection{Molecular gas properties of cluster galaxies at $0.2\lesssim z\lesssim2.0$}
We compare the results found for our galaxies in terms of molecular gas content, SFR, and depletion time with those found by other studies, including recent ones,  of both MS field galaxies and distant cluster galaxies, over a broad redshift range ${0.2\lesssim z\lesssim2.0}$, corresponding to ${\sim7.8}$~Gyr of cosmic time. 

Similarly to \citet{Noble2017}, we exploit recent observations of CO in cluster galaxies to study the evolution of the molecular gas content. We include CO detections of cluster galaxies at $z\sim0.2$ \citep{Cybulski2016}, $z\sim0.4-0.5$ \citep{Geach2011,Jablonka2013}; $z\sim1.1$ \citep{Wagg2012}, and $z\sim1.5-1.6$ \citep{Aravena2012,Noble2017}. 
We also include more recent results from { \citet{Rudnick2017}, \citet{Webb2017}, and \citet{Hayashi2018}} at $z\sim1.5-1.7$ { and those from \citet{Coogan2018} at $z\simeq2.0$,} as well as our CO(2-1) detection. This yields { 55 sources over 15 clusters}, including our detection, counted here as a  single source, with available estimates of the SFR and both stellar and total gas masses. Among the sources, we include two pairs of unresolved galaxies from \citet{Noble2017}, namely J0225-3680/3624 and  J0225-396/424. Consistently with \citet{Noble2017}, we count them as single sources and adopt physical quantities reported in their Table~1 for each of the two pairs. { We note that the work by \citet{Hayashi2018}, published after the submission of the present paper, reports 17 galaxies detected in CO belonging to the cluster XMMXCS~J2215.9-1738 at $z=1.5$. Their work includes CO results from \citet{Hayashi2017}. The sample of \citet{Stach2017} comprising six sources with independent CO detections from the same cluster is also entirely included in \citet{Hayashi2018}. Therefore, concerning XMMXCS~J2215.9-1738 cluster galaxies, we consider their  more recent results.}

\begin{figure}[]\centering
\subfloat{\includegraphics[width=0.53\textwidth]{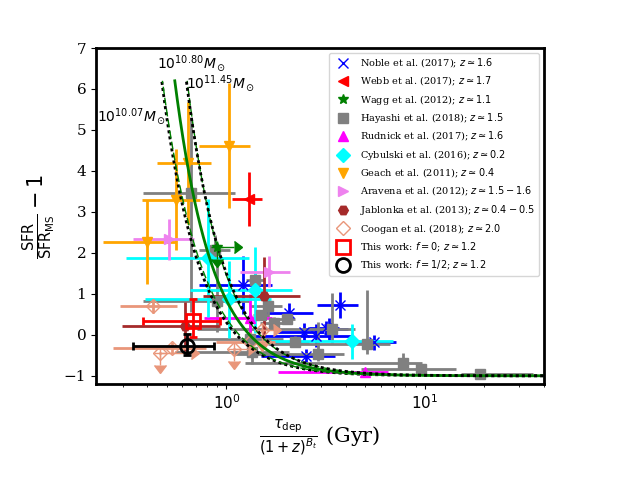}}
\subfloat{\includegraphics[width=0.53\textwidth]{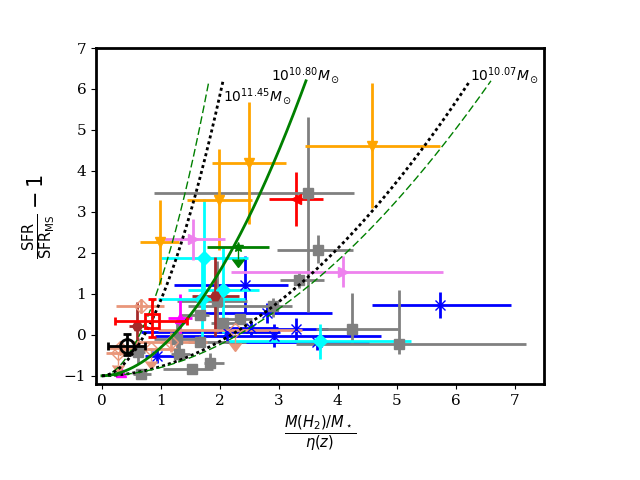}}\\
\caption{Fractional offset from the star forming { MS}  as a function of the molecular gas depletion timescale (left) and { molecular gas to stellar mass ratio} (right). Cluster galaxies at $0.2\lesssim z\lesssim2.0$ detected at CO are shown. 
{Data points from this work refer to the cases  where i) the two blended target sources equally contribute to the observed CO(2-1) flux ($f=1/2$) or ii) only one of the two contributes to the observed flux ($f=0$)}. 
In both panels the solid green curve is the scaling relation for field galaxies found by \citet{Tacconi2018} for galaxies with ${\rm Log}(M_\star/M_\odot)$=10.8 {and with an effective radius equal to the mean value found by \citet{vanderWel2014}  for star forming galaxies for given $z$ and $M_\star$}. The {green} dashed lines show the statistical 1-$\sigma$ uncertainties in the model. { The dotted black lines show the same scaling relation as the solid green lines, but for different stellar masses ${\rm Log}(M/M_\star)=$~10.07 and 11.45, that correspond to the stellar mass range spanned by the data points.}}
\label{fig:gas_properties}
\subfloat{\includegraphics[width=0.5\textwidth]{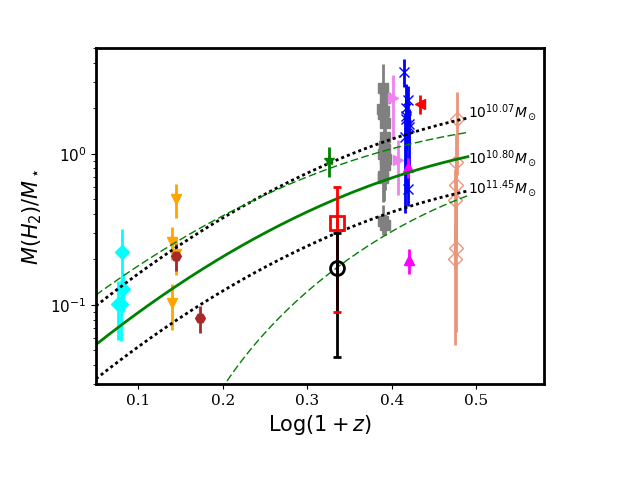}}\\
\caption{Evolution of the {molecular gas to stellar mass ratio} as a function of the redshift for cluster galaxies at $0.2\lesssim z\lesssim2.0$ detected at CO. Color code is analogous to that of Fig.~\ref{fig:gas_properties}. 
The solid { green} curve is the scaling relation found by \citet{Tacconi2018} for field galaxies in the main sequence and  with ${\rm Log}(M_\star/M_\odot)$=10.8, { where an effective radius equal to the mean value found by \citet{vanderWel2014} for star forming galaxies for given $z$ and $M_\star$ is assumed.} The {green} dashed lines show the statistical 1-$\sigma$ uncertainties in the model. { The dotted black lines show the same scaling relation as the solid green line, but for different stellar masses ${\rm Log}(M/M_\star)=$~10.07 and 11.45, that correspond to the stellar mass range spanned by the data points.} }
\label{fig:gas_fraction}
\end{figure}

In Fig.~\ref{fig:gas_properties}, we show the fractional offset from the star-forming MS as a function of the molecular gas depletion timescale (left) and { molecular gas to stellar mass ratio} (right), where we limit ourselves to galaxies with estimated stellar mass $>10^{10}~M_\odot$. For each galaxy, we also require SFR$<6$~SFR$_{\rm MS}$, where ${\rm SFR}_{\rm MS}$ is the SFR estimated using the{ \citet{Speagle2014}} model for MS field galaxies of stellar mass and redshift equal to those of the galaxy considered. This selection yields{ 49 sources over the 15 clusters}, shown in Fig.~\ref{fig:gas_properties}. In Fig.~\ref{fig:gas_fraction}, we show the evolution of the {molecular gas to stellar mass ratio} for the same{ 49} cluster galaxies.
{ In the left and right panels of Fig.~\ref{fig:gas_properties}, the x-axis values $\tau_{\rm dep}$ and $M({\rm H}_2)/M_\star$ are rescaled by $(1+z)^{\rm B_t}$ and $\eta(z)$, respectively, to take the redshift dependence inferred by \citet{Tacconi2018} into account, where B$_{\rm t}=-0.62$ and ${\rm Log}\,\eta(z) = -3.62\times[{\rm Log}(1+z)-0.66]^2$. }
When needed, we corrected the molecular mass estimates from the literature to take the different conversion factors $\alpha_{\rm CO}$ into account. We assume $\alpha_{\rm CO}=4.36~M_\odot({\rm K~km}~{\rm s}^{-1}~{\rm pc}^2)^{-1}$,
consistently with the value adopted throughout this work, while for each galaxy we assume the excitation level adopted in the corresponding work.
We stress that by assuming the same $\alpha_{\rm CO}$ for all galaxies, we only aim at having comparable molecular gas content for the galaxies considered; according to our selection, the majority of them lie around the MS, which justifies our choice for $\alpha_{\rm CO}$, as further described in the following.

{ \citet{Coogan2018} reported CO line fluxes for eight cluster galaxies, six of them also have stellar mass estimates and are considered in this work. All six sources are detected in CO(4-3) while two of them are also detected in CO(1-0). To estimate the molecular gas mass, we used the CO(1-0) flux, when available. Alternatively, we used the CO(4-3) flux by assuming an excitation level $r_{41}= L^{\prime}_{CO(4-3)}/L^{\prime}_{CO(1-0)}=0.36$ equal to the mean value inferred from the two galaxies with both CO(1-0) and CO(4-3) detections.}

As noted by \citet{Noble2017}, including galaxies significantly above the MS might lead to biased-high molecular gas to stellar mass ratios. Such sources are also typically associated with lower values for $\alpha_{\rm CO}$. Among the {  49} sources considered in Figs.~\ref{fig:gas_properties} and \ref{fig:gas_fraction}, those that have SFR$>3$ x ~SFR$_{\rm MS}$ are mainly at $z<0.6$, with the only exceptions represented by the galaxy at $z\simeq1.7$ of \citet{Webb2017}, { the galaxy with ID number 51858 of \citet{Aravena2012}, at $z\simeq1.6$, the sources  ALMA.12 and ALMA.17 at $z\simeq1.5$ of \citet{Hayashi2018} (see their Table~2)},  and the source at $z\simeq1.1$ of \citet{Wagg2012}. Concerning the last one, we note that their estimated SFR$<150~M_\odot/$yr is an upper limit due to the likely AGN contamination \citep{Wagg2012}. For the sake of completeness we do not reject these galaxies. This choice will not affect our conclusions.

{In Figs.~\ref{fig:gas_properties} and \ref{fig:gas_fraction}, the uncertainties of $\tau_{\rm dep}$ and 
$M({\rm H}_2)/M_\star$ for the sources with CO detections from the literature are estimated considering $M_\star$, $M({\rm H_2})$, and SFR as independent variables. 
In particular, we checked that the SFRs from the literature{ used in this work} are indeed independent from the associated CO observations and are based on spectral energy distribution (SED) fitting \citep{Aravena2012,Noble2017,Rudnick2017}, infrared luminosity \citep{Wagg2012,Jablonka2013,Cybulski2016,Stach2017,Webb2017}, and polycyclic aromatic hydrocarbon 7.7~$\mu$m emission \citep{Geach2011}. }
{ \citet{Hayashi2018} provide SFR estimates obtained using either i) SED fitting to the optical-to-mid-infrared photometry or ii) a combination of ultraviolet and infrared luminosities.
We adopted the latter when both SFR estimates are available for a given source of their sample, analogously to what has been done by the authors. However, we stress that for the six sources of \citet{Stach2017} included in \citet{Hayashi2018}, the SFRs from the two studies are discrepant up to a factor of $\sim5$ in some cases, which is due to the high uncertainties typically associated  with SFR estimates (see also Sect.~\ref{sec:discussion}). }
{ \citet{Coogan2018} report SFR estimates derived from  the CO(4-3), 1.4~GHz, or 870$\mu$m  fluxes. We adopted the SFR estimates based on the 870$\mu$m  fluxes.}

\section{Discussion}\label{sec:discussion}
\subsection{Main sequence, star formation, and depletion time}
Our results based on CO(2-1) observations show that the two unresolved galaxies have both SFR and sSFR consistent with those of MS galaxies; see, for example, our Fig.~\ref{fig:gas_properties}, and Fig.~7 of \citet[][]{Zeimann2013}. These conclusions are valid for both presented scenarios where either $f=0$ or $f=1/2$, {  and where the ${\rm SFR(24\mu m)}$ is adopted}. Our results are also { fairly} consistent with those found by considering instead the ${\rm SFR(H\alpha)}$ from \citet{Zeimann2013}, given the large uncertainties associated with the physical quantities such as SFR, $M_\star$, and $M({\rm H_2})$; see Tables~\ref{tab:cluster_galaxies_properties} and \ref{tab:NOEMAresults}.

In the case where $f=1/2$, that is, where both sources contribute equally to the observed CO(2-1) flux,  the SFR(24$\mu$m) is $\sim1/5$~SFR(H$\alpha$), for each of the two galaxies, separately. 
SFR estimates can be in fact uncertain by a factor of $\sim2$ or higher \citep[e.g.,][]{Kobayashi2013}. 
This is mainly due to the scatter between the SFR and SFR indicators such as
the H$\alpha$ luminosity and to the specific physical assumptions such as those adopted in this work. For our two target sources, the discrepancy between SFR(24$\mu$m) and SFR(H$\alpha$) is still fairly compatible with the estimated uncertainties.  However, even if no AGN contribution was found by previous work \citep[e.g.,][]{Zeimann2013}, as noted in Sect.~\ref{sec:results}, some AGN contamination might occur, resulting in biased-high SFR(H$\alpha$) estimates. Such a scenario could favor a better agreement between the SFR(H$\alpha$) and the SFR(24$\mu$m).

{ The depletion time of our target galaxies ${\tau_{\rm dep}\sim0.4}$~Gyr is marginally  consistent with, but shorter
than  that of MS galaxies.}
{In fact, the model of{ \citet{Tacconi2018}} predicts, for MS field galaxies, a higher depletion time than that estimated for our target galaxies. The value ${\tau_{\rm dep}\simeq(0.8\pm0.1)}$~Gyr is found for MS field galaxies of stellar mass and redshift similar to those of our target sources (see Fig.~\ref{fig:gas_properties}).} 
The uncertainties associated with the adopted physical assumptions may explain such a $\tau_{\rm dep}$ discrepancy. In particular, we assumed fully excited gas. However, if the excitation level is lower, that is, $r_{21}<1$, we could have higher values of $M({\rm H}_2)$ { and, consequently, a higher $\tau_{\rm dep}$.}


A number of observational facts privilege the scenario where both target galaxies contribute to the observed CO(2-1) emission implying that the case where $f=1/2$ might be more realistic than $f=0$.  The two galaxies have in fact similar stellar masses and H$\alpha$-based SFRs, as found in previous work \citep{Zeimann2013}.  Furthermore the FWHM of the detected CO(2-1) line is $\sim700$~km/s which is large when compared to the typical width of $\sim$200-300~km/s found for cluster galaxies at similar redshift \citep[e.g.,][and references therein]{Noble2017,Rudnick2017}. Such a discrepancy suggests that the two galaxies might have a large relative velocity of $\sim$400~km/s and that the observed CO(2-1) emission is indeed coming from both target galaxies.

The fact that the two galaxies have a projected angular separation of 6~kpc and are not resolved in velocity by their CO(2-1) spectrum might also suggest that the two sources are in a pair. However, cluster galaxies can have large peculiar velocities, up to $\sim$2,000~km/s for rich clusters \citep[e.g.,][]{Sheth_Diaferio2001,Biviano2013}. Therefore we cannot exclude the possibly that  projection effects occur and that the 3D physical separation between the two target galaxies is large. 

Since the two sources have molecular gas content and sSFR compatible with those of MS galaxies we suggest that they are not yet interacting. Alternatively, under the assumption that they are in a pair, their interaction has not yet led to their starburst phase. 
Since the two galaxies are not resolved by our NOEMA observations, we  cannot distinguish between different gas processing mechanisms (e.g., gas stripping). 

\subsection{Future perspectives}
Higher-resolution and higher-frequency observations at millimeter wavelengths will enable us to deblend the two sources both in velocity and in projected space. 
The presence/absence of asymmetries or disturbed morphology could be used to distinguish between different gas processing mechanisms. The spatial distribution of the CO emission, in particular, will allow us to understand if CO is confined within each galaxy or if it is associated with a more extended emission connecting the two galaxies. \textcolor{black}{If the latter scenario is confirmed, the two sources may be interacting. Such interaction may lead to a starburst phase by $z\sim0$, as observed for example in local major merging gas-rich pairs such as the Antennae Galaxies \citep{Gao2001,Ueda2012,Whitmore2014}.} 

Higher-resolution observations of higher-J CO transitions such as CO(4-3) could also be used to better estimate both the continuum flux and the dust mass in the Rayleigh-Jeans regime, as well as probe the excitation level, in combination with the CO(2-1) detection presented in this work. 




\subsection{Comparison with other distant cluster galaxies}
The data available from the literature on the SFR and depletion time of $z>1$  cluster galaxies (Fig.~\ref{fig:gas_properties}) are in overall agreement with the model of field galaxies. Similarly, the { molecular gas to stellar mass ratio} (Fig.~\ref{fig:gas_fraction}) is fairly consistent with model predictions of MS galaxies in the field, given the dispersion in data points and the uncertainties underlying the physical assumptions such as those of the $\alpha_{CO}$ conversion factor, and of both SFR and stellar mass estimates. However, as also noted in \citet{Noble2017} on the basis of their CO observations and those prior to 2017, the{ molecular gas to stellar mass ratio} of $z>1$ cluster galaxies from the literature seems to be generally higher than that of field galaxies. This statement does not apply to either our galaxies or recent CO observations of cluster galaxies at $z\sim1.6$ and ${z\sim2.0}$ from \citet{Rudnick2017} { and \citet{Coogan2018}}, respectively. Increasing the sample of distant CO detected cluster galaxies will certainly help to achieve a better understanding of their statistical molecular gas properties. 

{ The fact that the two target galaxies have molecular gas mass and depletion times that are marginally compatible with but smaller than those of MS field galaxies suggests that the molecular gas has not been adequately refueled. We speculate that the cluster environment might have played a role in preventing the refueling via environmental mechanisms such as galaxy harassment, strangulation, ram-pressure, or tidal stripping.  }


\subsection{Additional comments on physical assumptions and models}
Some additional comments are needed concerning the physical assumptions and conventions adopted in this work. 
The exact value of the H$_2$-to-CO conversion factor is unknown. By assuming a lower $\alpha_{CO}\simeq(1-2)~M_\odot({\rm K~km}~{\rm s}^{-1}~{\rm pc}^2)^{-1}$ \citep{Wagg2012,Stach2017,Webb2017}, typical of (ultra-) luminous infrared galaxies \citep[][]{Bolatto2013}, we find shorter depletion time scales and smaller{ molecular gas to stellar mass ratios} for the two selected galaxies, which imply a stronger discrepancy with respect to the values found for MS field galaxies. These findings further support our choice for $\alpha_{CO}$.  Furthermore, as noted above, if the excitation level were found to be lower than assumed, that is, $r_{21}<1$, we would obtain higher {molecular gas to stellar mass ratios} and higher depletion time scales, leading to a better agreement with respect to the model for MS galaxies.

We also stress here that the stellar mass estimates used in this work to infer quantities such as $M({\rm H}_2)/M_\star$ and sSFR  rely on stellar population synthesis models and have statistical uncertainties $\sim(0.10-0.14)$~dex \citep[e.g.,][]{Roediger_Courteau2015}. Additional $\sim$0.25~dex uncertainty may be added because of the unknown initial mass function \citep[][and references therein]{Wright2017}. This yields a typical uncertainty of $\sim0.3$~dex on the stellar mass. Our two target galaxies have stellar masses with estimated uncertainties that are even higher and equal to $\sim$0.5~dex; see Table~\ref{tab:cluster_galaxies_properties}. { Stellar masses of our two target galaxies have been estimated by \citet{Zeimann2013} using a \citet{Chabrier2003} initial mass function.}



{ We note that both the depletion time and the molecular gas to stellar mass ratio only weakly depend on the effective radius R$_{\rm e}$. \citet{Tacconi2018} found that both $\tau_{\rm dep}$ and  $M({\rm H_2})/M_\star$ scale as $\propto({\rm R_e/R_{\rm e0}})^{0.11}$, where R$_{\rm e0}=8.9$~kpc~$(1+z)^{-0.75}[M_\star/(5\times10^{10}M_\star)]^{0.22}$ is the mean effective radius of star-forming galaxies as a function of $z$ and $M_\star$ from \citet{vanderWel2014}. For our two target sources, we have 
R$_{\rm e}\simeq$~3-4~kpc and R$_{\rm e0}=5.2$~kpc, which implies a negligible correction of $\lesssim5$\% on the model values of $\tau_{\rm dep}$ and $M({\rm H_2})/M_\star$, that are reported in Figs.~\ref{fig:gas_properties} and \ref{fig:gas_fraction} 
in the case of ${\rm R_e}={\rm R_{e0}}$.}

{ We also checked that our results are fairly independent of the specific model used for MS field galaxies. We adopted the model prescriptions described by \citet{Tacconi2018} whose study supersedes the previous one by \citet{Genzel2015}. 
In particular, we used the \citet{Speagle2014} model for the ${\rm SFR_{MS}}$, also used by \citet{Tacconi2018}. Adopting instead the model by \citet{Whitaker2012} used by \citet{Genzel2015} would lead to a similar ${\rm SFR_{MS}}$, only slightly higher  by  $(19\pm20)\%$ on average, well within the typical SFR uncertainties, where the reported value is the median value estimated from the data points reported in Fig.~\ref{fig:gas_properties} and the uncertainty is the rms around the median.
If we limit ourselves to galaxies with redshift and stellar mass equal to those of our target galaxies and if we use the \citet{Genzel2015} model prescriptions for star-forming galaxies $\tau_{\rm dep}$ and $M({\rm H}_2)/M_\star$ are $\sim20\%$ higher than those obtained using the model prescriptions by \citet{Tacconi2018}. The \citet{Genzel2015} and \citet{Tacconi2018} models are therefore compatible with each other within the associated uncertainties.  We refer to \citet{Tacconi2018} for a discussion.}

\section{Conclusions}\label{sec:conclusions}
In this work, we report a CO(2-1) detection (${\rm S/N}=5.4$) from two unresolved cluster galaxies at $z=1.2,$ obtained with
the IRAM NOEMA interferometer. The two galaxies are separated by 6~kpc in the projected space and are located at a projected radial distance of 500~kpc from the center of the distant cluster ISCS J1426.5+3339.

The two sources are selected as the ones with the highest  SFR(H$\alpha$), that is, $\gtrsim120$~$M_\odot$/yr, within the \citet{Zeimann2013} catalog of $1.0 < z < 1.5$ cluster galaxies. The galaxies in this catalog belong to a total of 18 clusters from the Irac Shallow Cluster Survey  \citep[ISCS,][]{Eisenhardt2004,Brodwin2006}.
The halo mass of the cluster ISCS~J1426.5+3339, {which the two selected galaxies belong to,} is expected to be in the range  $\sim(0.8-2)\times10^{14}~M_\odot$, corresponding to a virial radius of $\sim1$~Mpc \citep{Wagner2015}.

The CO detection presented in this work is the first detection resulting from a wider search of molecular gas in distant cluster galaxies. It  is also  the first one from $1 < z < 1.4$ cluster galaxies with no  reported{ clear} evidence of AGNs. 

Our observations yield an improvement, by a factor of ten in accuracy, of an unknown weighted mean of the redshifts of the two sources, $z=1.163\pm0.001$, based on the CO(2-1) emission line peak. {  Our redshift estimate is also consistent with those obtained for the two target galaxies using the HST WFC3 camera and the near-infrared grism G141 \citep{Zeimann2013}.} 

The integrated CO(2-1) line luminosity implies a total molecular gas mass $M({\rm H_2})=(2.2^{+0.5}_{-0.4})\times10^{10}$~$M_\odot$ for the two blended sources. A Galactic CO-to-H$_2$ conversion factor $\alpha_{CO}=4.36~M_\odot\,({\rm K~km~s}^{-1}~{\rm pc}^2)^{-1}$  is used for the calculation. We also assumed fully excited gas. 

The FWHM of the detected CO(2-1) line is $\sim700$~km/s which is large when compared to the typical width of $\sim$200-300~km/s found for cluster galaxies at similar redshift \citep[e.g.,][and references therein]{Noble2017,Rudnick2017}. Such discrepancy suggests that the two galaxies might have a large relative velocity of $\sim$400~km/s and that the observed CO emission indeed originates from both galaxies. However, since the two galaxies are not resolved by our NOEMA observations we  cannot distinguish between different gas processing mechanisms (e.g., gas stripping).

{ By assuming that both blended sources contribute equally to the observed CO(2-1) and 24$\mu$m (Spitzer MIPS) emission, we estimate the ${\rm SFR(24\mu m)}=(28^{+12}_{-8})~M_\odot$/yr, the depletion timescale 
$\tau_{\rm dep}=(3.9^{+1.4}_{-1.8})\times10^8$~yr, } and the molecular gas to stellar mass ratio
$M({\rm H_2})/M_\star=0.17\pm0.13$, for each of the two galaxies.

{ The SFR(24$\mu$m) is consistent with that inferred from standard scaling relations involving molecular gas content \citep[e.g.,][]{Daddi2010} and is $\sim$1/5~SFR(H$\alpha$).} This result suggests that the SFR(H$\alpha$) might be biased-high, possibly due to some AGN contamination. Nevertheless, the two SFR estimates are { marginally} consistent with each other within their uncertainties. 
{  The depletion time of our target galaxies is marginally  consistent with but shorter than that of MS galaxies} of stellar mass and redshift similar to those of our target sources, ${\tau_{\rm dep}\simeq(0.8\pm0.1)}$~Gyr \citep{Tacconi2018}.

Nevertheless, given the uncertainties associated with both model predictions and observations, { the SFR, the molecular gas mass, and the depletion time of our target galaxies} are in overall agreement with those found for other distant cluster galaxies at similar redshifts and with the model from{ \citet{Tacconi2018}} for  MS field galaxies of similar mass and redshift of our two sources. 
These results suggest that the two galaxies are not interacting, but are possibly close to interacting. Alternatively, their interaction has not led to their starburst phase yet. Such a starburst phase may be reached by $z\sim0$, as observed, for example, in local major merging gas-rich pairs such as the Antennae Galaxies.


{ However, the fact that the two target galaxies have molecular gas mass and depletion time that are marginally compatible with but smaller than those of MS field galaxies suggests that the molecular gas has not been adequately refueled. We speculate that the cluster environment might have played a role in preventing the refueling via environmental mechanisms such as galaxy harassment, strangulation, ram-pressure, or tidal stripping. }

Higher-resolution and higher-J CO observations at millimeter wavelengths will allow us to spatially resolve the two sources and possibly distinguish between different gas processing mechanisms. 

\begin{acknowledgements}
This work is based on observations carried out under project number S17BJ with the IRAM NOEMA Interferometer. IRAM is supported by INSU/CNRS (France), MPG (Germany) and IGN (Spain). GC thanks Cinthya Herrera (LOC at NOEMA, Grenoble) and all IRAM staff for the hospitality in Grenoble and precious help concerning data reduction. { We thank the anonymous referee for helpful comments that led to a substantial improvement of the paper. GC thanks Paola Dimauro for the help concerning the use of SExtractor. CC acknowledges funding from the European Union's Horizon 2020 research and innovation program under the Marie Sklodowska-Curie grant agreement No. 664931.}
\end{acknowledgements}


\begin{table}[tb]\centering
\begin{tabular}{ccccccccc}
\hline\hline
cluster ID & galaxy ID & $z_{spec}$ & Log($M_\star/M_\odot$)  & {Flux  (H$\alpha$+[NII])} & SFR(H$\alpha$) & sSFR(H$\alpha$) & $\tau_{\rm dep}{\rm (H\alpha)}$ & R$_{\rm e}$ \\
   & & & & {\tiny ($10^{-17}\frac{erg}{s~cm^2}$)} &  ($M_\odot/{\rm yr}$) & (Gyr$^{-1})$ & ($10^8$~yr) & (kpc) \\ 
 (1) & (2) & (3) & (4) & (5) & (6) & (7) & (8) & (9) \\
 \hline
 { ISCSJ1426.5+3339} & { J142626.1+333827} & 1.17$\pm$0.01  & $10.8\pm0.5$ & 79.1$\pm$5.3 & $127^{+100}_{-75}$ & $2.0^{+2.1}_{-1.9}$ & $0.87^{+0.55}_{-0.70}$ & 3.0  \\ 
 & { J142626.1+333826} & 1.17$\pm$0.01 & $10.8\pm0.5$ &  83.7$\pm$4.8 & $134^{+100}_{-75}$ & $2.1^{+2.2}_{-2.0}$ &  $0.82^{+0.49}_{-0.63}$ & 3.9 \\ 
 \end{tabular}
\caption{Properties of the targets. (1) cluster and (2) galaxy name; (3) spectroscopic (HST WFC3) redshift of the galaxy; (4) galaxy stellar mass; (5) H$\alpha$+[NII] line flux; (6) SFR(H$\alpha$) estimated following \citet{Zeimann2013}; (7)  sSFR(H$\alpha$)=SFR(H$\alpha)/M_\star$; (8) $\tau_{\rm dep}{\rm (H\alpha)}=M({\rm H_2})/{\rm SFR(H}\alpha)$ estimated assuming that the two sources equally contribute to the observed CO(2-1) flux, i.e., $f=1/2$; {(9) effective radius estimated using SExtractor and the HST WFC3 image.}  Columns (3-5) are from \citet{Zeimann2013}.}
\label{tab:cluster_galaxies_properties}
\vspace{1cm}
\begin{tabular}{cc|l}
\hline\hline
(1) & galaxy ID & J142626.1+333827~/~J142626.1+333826 \\
\hline
(2)$^\ast$ & $z$ & $1.163\pm0.001$, from CO(2-1) \\
\hline
(3)$^\ast$ &    FWHM & $(733\pm200)$~km/s, from CO(2-1)\\
\hline
(4)$^\ast$ &    $S_{CO(2-1)}$ & $(0.27^{+0.06}_{-0.05})$~Jy~km/s \\
\hline
(5)$^\ast$ &    $L^{\prime}_{CO(2-1)}$ & $(4.9^{+1.1}_{-0.9})\times10^9$~pc$^2$~K~km~s$^{-1}$\\ 
\hline
(6)$^\ast$ & $M({\rm H_2})$ & $(2.2^{+0.5}_{-0.4})\times10^{10}$~$M_\odot$\\
\hline
(7)$^\ast$ & continuum  & $<33$~$\mu$Jy\\
\hline
(8)$^\ast$ & $M_{\rm dust}$ & $<4.2\times10^8$~$M_\odot$\\
\hline
(9) & { SFR(24~$\mu$m)} &  ${(52^{+21}_{-15})}$~$M_\odot/$yr~$\,(f=0)$; ${(28^{+12}_{-8})}$~$M_\odot/$yr~$\,(f=1/2)$\\
\hline
(10) & { sSFR(24~$\mu$m)} & ${(0.82^{+0.65}_{-0.64}}$~Gyr$^{-1}$~$\,(f=0)$; ${(0.44^{+0.36}_{-0.34})}$~Gyr$^{-1}$~$\,(f=1/2)$\\
\hline
(11) & $\tau_{\rm dep}$ & ${(4.2^{+1.6}_{-1.9})\times10^8}$~yr~$\,(f=0)$ ; ${(3.9^{+1.4}_{-1.8})\times10^8}$~yr~$\,(f=1/2)$ \\
\hline
(12) & $M({\rm H_2})/M_\star$ & ${0.35^{+0.25}_{-0.26}}$~$\,(f=0)$ ; ${0.17\pm0.13}$~$\,(f=1/2)$\\
\hline
\end{tabular}
\caption{Summary of the results. Quantities marked with an asterisk (2-8) are estimated for the two blended sources. Quantities (9-12) refer to each of the two sources, separately. The two scenarios where both sources contribute equally to the observed CO(2-1) flux ($f=1/2$ ) or only one of the two does ($f=0$) are considered.}
\label{tab:NOEMAresults}
\end{table}

\end{document}